\documentclass[12pt]{article}

\usepackage{newtxtext,newtxmath}
\usepackage{graphicx}

\usepackage[letterpaper,margin=1in]{geometry}

\renewenvironment{abstract}
	{\quotation}
	{\endquotation}

\date{}

\makeatletter
\renewcommand{\fnum@figure}{\textbf{Figure \thefigure}}
\renewcommand{\fnum@table}{\textbf{Table \thetable}}
\makeatother

\usepackage{scicite}

\usepackage{url}

\def\scititle{
	A millisecond integrated quantum memory for photonic qubits
}
\title{\bfseries \boldmath \scititle}

\author{
	Yu-Ping~Liu$^{1,2,3}$, %\ast\dagger
	Zhong-Wen~Ou$^{1,2,3}$, 
	Tian-Xiang~Zhu$^{1,2,3}$, \and
	Ming-Xu~Su$^{1,2,3}$, 
	Chao~Liu$^{1,2,3}$, 
	Yong-Jian Han$^{1,2,3,4}$, \and
 Zong-Quan~Zhou$^{\mathrm{a},1,2,3,4}$, 
	Chuan-Feng~Li$^{\mathrm{b},1,2,3,4}$, 
	Guang-Can~Guo$^{1,2,3,4}$\and
	\small$^{1}$CAS Key Laboratory of Quantum Information,\and \small University of Science and Technology of China, Hefei, 230026, China.\and
	\small$^{2}$Anhui Province Key Laboratory of Quantum Network,\and \small  University of Science and Technology of China, Hefei, 230026, China.\and
	\small$^{3}$CAS Center for Excellence in Quantum Information and Quantum Physics,\and \small  University of Science and Technology of China, Hefei, 230026, China.\and
	\small$^{4}$Hefei National Laboratory, University of Science and Technology of China, Hefei, 230088, China.\and
	  \small$^\mathrm{a}$zq\_zhou@ustc.edu.cn\and
        \small$^\mathrm{b}$cfli@ustc.edu.cn\and
}

\begin{document} 

% Insert the title and author list
\maketitle

% Abstract, in bold
% There are strict length limits, and not all formats have abstracts.
% Consult the journal instructions to authors for details.
% Do not cite any references in the abstract.
\begin{abstract} \bfseries \boldmath
    Quantum memories for light are essential building blocks for quantum repeaters and quantum networks. Integrated operations of quantum memories could enable scalable application with low-power consumption.
    However, the photonic quantum storage lifetime in integrated optical waveguide has so far been limited to tens of microseconds, falling short of the requirements for practical applications. Here, we demonstrate quantum storage of photonic qubits for 1.021 ms based on a laser-written optical waveguide fabricated in a $\mathrm{^{151}Eu^{3+}}$:$\mathrm{Y_2SiO_5}$ crystal. Spin dephasing of $\mathrm{^{151}Eu^{3+}}$ is mitigated through dynamical decoupling applied via on-chip electric waveguides and we obtain a storage efficiency of $12.0\pm0.5$$\%$ at 1.021 ms, which is a demonstration of integrated quantum memories that outperforms the efficiency of a simple fiber delay line. Such long-lived waveguide-based quantum memory could support applications in quantum repeaters, and further combination with critical magnetic fields could enable potential application as transportable quantum memories.
\end{abstract}

\subsection*{Teaser}
    Photonic qubits are stored in an integrated optical waveguide for 1 ms with an efficiency outperforming a fiber delay line.

% The first paragraph of any Science paper does NOT have a heading
% Nor is it indented
\subsection*{Introduction}
\noindent
    Photonic quantum memory is a key enabler for constructing large-scale quantum networks \cite{qr0,dlcz,RN3,oqm} through approaches such as quantum repeaters \cite{qr0,RN3,space,sat} and transportable quantum memories \cite{sat1,sat2,RN47,10h,RN95}, which have been demonstrated in various atomic platforms \cite{RN170,DD,hanson,pan,lukin,ions,ion1,atom,bell,RN6,eit,qnc}. As an ensemble based quantum memory \cite{RN50,RN168,solid,RN165}, rare-earth-ion-doped crystals (REICs) have attracted much attention due to their long coherence lifetimes \cite{prb1,np,1min,10h,RN95,RN170} and wide bandwidths \cite{RN50,RN4,erfiber}. As a solid-state platform, integrated operations have been demonstrated with various fabrication techniques in REICs \cite{RN4,RN147,NSR,RN157,piqm}, enabling quantum storage in optical waveguide with high fidelity \cite{RN61,RN162}, large memory capacity \cite{m3,lnmul} and direct telecom interface \cite{RN195,ln3}. However, the storage lifetimes of these integrated devices have been constrained to tens of microseconds \cite{NSR,RN157} and the storage efficiency fall below that of a simple fiber delay line, posing substantial challenges for practical applications in quantum repeaters and transportable quantum memories.

    In this work, we demonstrate quantum memory for light with a $1/e$ lifetime of 1.9 ms, utilizing an integrated laser-written optical waveguide fabricated in a $\mathrm{^{151}Eu^{3+}}$:$\mathrm{Y_2SiO_5}$ crystal. Leveraging the noiseless photon echo (NLPE) protocol \cite{RN137}, we achieve a quantum storage efficiency of $12.0\pm 0.5$\% at a storage time of 1.021 ms, far surpassing the corresponding efficiency of a fiber delay line operating in the 1550 nm C-band. A coplanar electric waveguide is employed to efficiently apply the dynamical decoupling (DD) sequence, mitigating spin dephasing of the ions inside the optical waveguide \cite{EW,RN170}. %Time-bin qubits encoded with single-photon-level inputs are stored and retrieved with a fidelity of $F=89.7\pm 1.5$\%.

    %demonstrating the reliability and the potential advantages for integrated solutions for quantum memories \cite{int,RN170,DD,RN162}.

%Figures and tables:
%These should be inserted at the end of the main text, as below (not in the middle of the text).
%Refer to them using e.g.~Figure~\ref{fig:example} (or Fig.~\ref{fig:example}) and Table~\ref{tab:example}.

%Citing references:
%Science uses a numeric citation system. Cite references by number e.g.~\cite{example}.
%The template will combine reference numbers automatically~\cite{example,example2},
%including ranges~\cite{example,example2, example_preprint}.
%Reference author names and years should be stated in the reference list, not in the text.
%If you want to add a comment, use the syntax [see \cite{example} for details].
% Not \cite[see][for details]{example}. Unfortunately that isn't compatible with scicite.sty

%the reference list that directs readers there, like this~\cite{methods}.

%\begin{equation}
%	x=\frac{-b\pm\sqrt{b^2-4ac}}{2a}.
%	\label{eq:example} % Use a logical label
%\end{equation}

%\noindent\ Do not indent text immediately after an equation.
%They can be referred back to as e.g.~Equation~\ref{eq:example}.

% Research Articles and Reviews split the text into sections using headings
% Use a short (up 6 words) descriptive phrase, not generic 'Results' or 'Conclusions'
% Most other formats do not have headings, see the journal instructions to authors for details
\subsection*{Results}
\subsubsection*{Experiment setup}
    We employ a 0.01$\%$ doped $\mathrm{^{151}Eu^{3+}}$:$\mathrm{Y_2SiO_5}$ crystal with a size of $\mathrm{5 \times 4 \times 17 \ mm^3}$ (D1 $\times$ D2 $\times$ b axes) as the substrate. This material is of particular interest due to its remarkable properties, including a spin coherence lifetime of 10 hours \cite{10h}, coherent light storage capability for up to 1 hour \cite{RN95}, and quantum storage durations in the millisecond range \cite{RN170,DD}.A low concentration of $\mathrm{^{151}Eu^{3+}}$ ions is chosen to minimize spin inhomogeneous broadening, which could otherwise limit the coherence lifetime under critical magnetic field conditions \cite{eusim,lpj}. An optical waveguide is fabricated along the $b$ axis using a femtosecond-laser micromachining system \cite{RN171,RN121,RN162,NSR}. To realize the long-lived quantum storage based on DD control on the spin transitions, we have fabricated a coplanar electric waveguide \cite{EW,cpw} on the top surface of the crystal via lift-off technique to apply the radio-frequency (RF) pulses. The magnetic field generated by the coplanar electric waveguide spatially matches well with the guide mode of the optical waveguide as illustrated in Figure~\ref{setup}.
    
    The optical waveguide comprises 20 tracks arranged along the b axis of $\mathrm{Y_2SiO_5}$ crystal as shown in Figure~\ref{setup}, which collectively form a circular structure to confine light \cite{RN171,RN121}. This structure can in principle support the transmission of arbitrary polarization modes \cite{RN162}. The absorption depth of the $\mathrm{{^7}F{_0} \rightarrow {^5}D{_0}}$ transition for Eu$^{3+}$ ions inside the waveguide is 2.09 and 0.38 for input light polarized along D1 axis and D2 axes of $\mathrm{Y_2SiO_5}$ crystal, respectively. The coplanar electric waveguide, comprising three electrodes arranged along the $b$ axis, is depicted in Figure~\ref{setup}B. The two outer electrodes serve as the ground, while the middle electrode acts as the signal electrode \cite{cpw2}. The electrode widths are optimized to produce a sufficiently uniform magnetic field for the ions within the optical waveguide. These electrodes are fabricated on the D2 $\times$ b surface to maximize the coupling of the magnetic field with the Eu$^{3+}$ hyperfine transitions compared to the configuration on the D1 $\times$ b surface.

    The experimental setup is depicted in Figure~\ref{setup}. The laser source (not shown) is a frequency-doubled, stabilized semiconductor laser with a center frequency of 516.847 THz and a linewidth of 0.4 kHz, resonant with $\mathrm{{^7}F{_0} \rightarrow {^5}D{_0}}$ transition of $^{151}$Eu$^{3+}$ ions (Figure~\ref{trace}A). The RF signal, resonant with the the hyperfine transition $|\pm1/2\rangle_g \rightarrow |\pm3/2\rangle_g$ at 34.5 MHz, is generated by an arbitrary waveform generator (Zurich HDAWG) and amplified by a 53-dB amplifier (Bruker RFA-0.1/250-150). The optical signal and control beams are generated using double-pass acousto-optic modulator (AOM). The control beam is in orthogonal polarization to that of the signal mode, allowing the PBS to effectively filter out coherent noise generated by the control pulses. Two single-passed AOM gates and a filter crystal are employed to reduce noise in the temporal and frequency domains, respectively. The filter crystal, a 0.1$\%$ doped $\mathrm{^{151}Eu^{3+}}$:$\mathrm{Y_2SiO_5}$ with a length of 30 mm, provides a typical absorption depth of 9.1 with a 2-MHz transparent window centered at the signal frequency. Both the memory crystal and the filter crystal are maintained at 3.2 K using a closed-cycle cryostat (Montana Instruments).
    
    The coplanar electric waveguide and the cables inside the cryostat together exhibit a reflection of $-25$ dB and a transmission of $-0.3$ dB at the operating frequency of 34.5 MHz. This integrated configuration achieves a Rabi frequency of $\Omega = 16.7$ kHz for the hyperfine transition $|\pm1/2\rangle_g \rightarrow |\pm3/2\rangle_g$ at an input peak power of 4 W (Figure~\ref{setup}C). The coplanar electric waveguide facilitates efficient spin driving with straightforward integration, substantially reducing the required RF power compared to conventional bulk RF coils \cite{RN95,RN170}. For comparison, we also apply an RF field using bulk coils within the same system. The coil, with a length matching that of the crystal, consists of 15 turns and has a radius of 10 mm. Achieving the same Rabi frequency requires a power input of 330 W, which is 82 times higher than that of the electric waveguide.
   The Rabi frequency can be further increased by narrowing the electrode width, albeit at the cost of increased RF field inhomogeneity. In this work, we select an electrode width of 150 $\upmu$m to strike a balance between minimizing the required RF power and maintaining acceptable field homogeneity. 

\subsubsection*{Storage of single-photon level inputs}
\begin{figure} 
	\centering
	\includegraphics[width=1\textwidth]{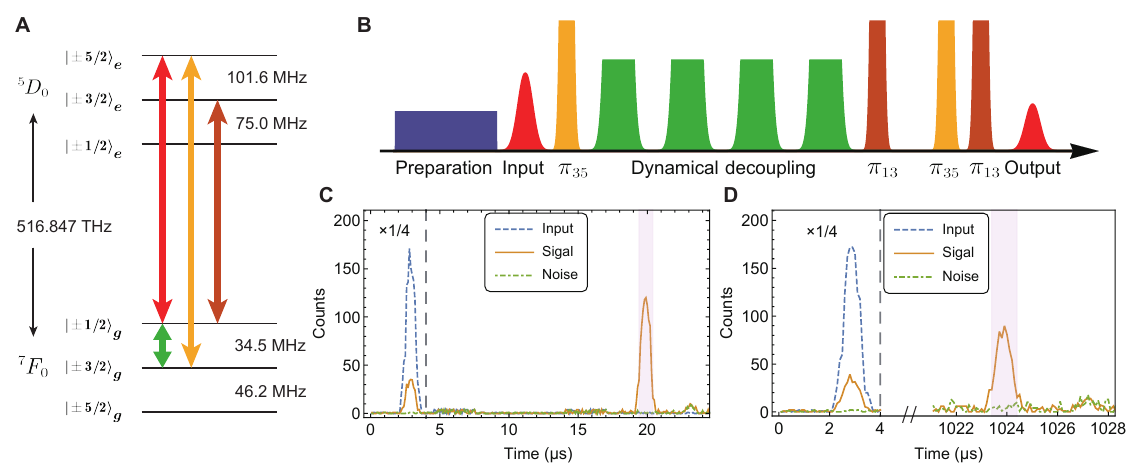} 
	\caption{\textbf{Integrated storage of single-photon level inputs for 1.021 ms.}
            (\textbf{A}) Energy level structure of the $\mathrm{{^7}F{_0}\rightarrow{^5}D{_0}}$ transition of $\mathrm{^{151}Eu^{3+}}$ ions in $\mathrm{Y_2SiO_5}$ crystals at zero magnetic field.
            (\textbf{B}) Time sequence for the NLPE quantum memory with DD. After initial spectral preparation, the input pulse is absorbed by inhomogeneously broadened $\mathrm{^{151}Eu^{3+}}$ ions. The first $\pi_{35}$ transfer the photonic excitation to spin-wave excitation. DD based on the XY4 sequence is employed rephase the spin transition. The remaining three optical $\pi$ pulses rephase the optical transition, and the final echo is retrieved at the same frequency as the input. The color of each pulse corresponds to the resonance transitions, as indicated in subfigure (A).
            (\textbf{C}) and (\textbf{D}) Photon-counting histograms for NLPE memory without (C) and with DD (D), respectively. The blue dashed line represents the input, while the orange solid line and the green dash-dotted line represent the signal and noise, respectively, with and without inputs. Data on the left side of the dashed gray line are scaled by a factor of 1/4 for visualization. The light purple shaded region highlights the 1.1 $\upmu$s detection window. With an average input photon number per pulse $\mu=1.07$, the signal-to-noise ratio (SNR) is $50.3\pm16.7$ for (C) and $13.1\pm2.4$ for (D).
        }
	\label{trace} 
\end{figure}

    The $\mathrm{{^7}F{_0} \rightarrow {^5}D{_0}}$ transition of $\mathrm{^{151}Eu^{3+}}$ is a strictly forbidden transition and becomes weakly allowed in $\mathrm{Y_2SiO_5}$ crystals. Given the low sample absorption depth ($\sim$1.5), we employ the NLPE protocol for spin-wave quantum storage. This approach can fully utilize the original sample absorption to capture input photons, making it particularly well-suited to the current sample \cite{RN137,RN143,NSR}. The experimental time sequence is illustrated in Figure~\ref{trace}B. An absorption band of 1.8 MHz is prepared within a 4-MHz transparent window on the $|\pm1/2\rangle_g \rightarrow |\pm5/2\rangle_e$ transition (figure~\ref{smfig15}B). The input signals are weak coherent pulses with an average photon number per pulse $\mu=1.07$. The Gaussian-shaped pulses have a FWHM of 0.8 $\upmu$s and a total duration of 1.7 $\upmu$s. Four optical $\pi$ pulses are required to implement the NLPE memory where $\pi_{13}$ pulses drive the transition $|\pm1/2\rangle_g \rightarrow |\pm3/2\rangle_e$ and $\pi_{35}$ pulses drive $|\pm3/2\rangle_g \rightarrow |\pm5/2\rangle_e$. We use adiabatic $\pi$ pulses with a bandwidth of 2.2 MHz and a duration of 1.5 us to drive the optical transition. The peak power is 50 mW for $\pi_{35}$ and 55 mW for $\pi_{13}$, respectively, corresponding to a same Rabi frequency of 1.2 MHz. In bulk demonstrations, achieving the same Rabi frequency would require six times more power \cite{RN137}. The final echo is emitted at the same frequency as that of the input signal \cite{NSR}. As shown in Figure~\ref{trace}C, the NLPE storage efficiency is $17.8\pm 0.9$$\%$ with a detection window of 1.1 $\upmu$s at a storage time of 17 $\upmu$s. The noise per temporal mode $p_N=0.38\pm 0.13\%$ and the SNR is $50.7\pm16.7$.
    
    When an input photon is stored as a collective spin-wave excitation of an atomic ensemble, the spin-wave excitation dephases within tens of $\upmu$s due to spin inhomogeneous broadening on the order of tens of kHz \cite{NSR}. In our system, the $1/e$ spin-wave storage lifetime of NLPE memory is approximately 42 $\upmu$s (figure~\ref{smfig2}B). To overcome this limit, we employ the DD to rephase the spin transitions and extend the storage lifetime to the spin coherence lifetime. Previously, this technique has only been demonstrated with atomic frequency comb (AFC) quantum memories in bulk crystals \cite{DD,RN170}. In our experiment, the DD sequence is applied between the first $\pi_{35}$ and the $\pi_{13}$ pulse, to rephase the spin transition between $|\pm1/2\rangle_g$ and $|\pm3/2\rangle_g$. We use the XY4 sequence (X-Y-X-Y), which is designed to preserve arbitrary initial states. Adiabatic pulses are employed in the sequence, offering better control over spin transitions. These pulses produce a flatter spectrum in the frequency domain compared to square pulses, ensuring consistent driving across the entire bandwidth \cite{DD,RN170}. The lifetime of the NLPE with DD (NLPE-DD) is extended to $1.90\pm0.04$ ms (figure~\ref{smfig3}C). At a storage time of 1.021~ms, the storage efficiency is $12.0\pm0.5$$\%$ for a detection window of 1.1~$\upmu$s  (Figure~\ref{trace}D). The noise per temporal mode $p_N=0.98\pm0.15\%$, resulting a SNR of $13.1\pm2.4$. 
    %By comparing this efficiency to the NLPE efficiency and considering the spin dephasing loss, we can estimate the spin rephasing efficiency of our DD sequence is 99.6\%.
    
    The noise notably increases after introducing DD into NLPE, primarily due to imperfections in the RF $\pi$ pulse within the DD sequence. Imperfect $\pi$  pulses reduce storage efficiency and leave residual population in $|\pm3/2\rangle_g$ after the DD process. Subsequently, the optical $\pi_{35}$ pulse transfers this residual population to $|\pm5/2\rangle_e$, resulting in indistinguishable spontaneous emission noise. By comparing the NLPE storage efficiencies with and without DD, we could estimate that the XY4 sequence leaves $0.3\pm0.2$\% population in $|\pm3/2\rangle_g$ in this device (see for details). This issue can be mitigated by extending the width of the middle electrode to improve the quality of DD pulses across the ensemble, albeit at the cost of increased heating (figure~\ref{smfig3}). Our implementation of NLPE-DD in a waveguide achieves an SNR approximately twice that of previous demonstrations of AFC memories with DD in bulk crystals \cite{DD,RN170}. The noise levels in both our experiments and previous studies are approximately the same, and the enhanced SNR in our work is primarily due to the improved storage efficiency of the NLPE protocol, despite a sample with lower concentration of Eu$^{3+}$ ions is used in the current work.

\subsubsection*{Storage of time-bin qubits}
    To benchmark the quantum storage performances of the current device, we further encode time-bin qubits on the input pulses. Four states are prepared as input qubits: $|e\rangle$, $|l\rangle$, $|e\rangle+|l\rangle$, and $|e\rangle+i|l\rangle$, where $|e\rangle$ and $|l\rangle$ represent the early bin and the late bin, respectively. Each bin has a duration of 1.7 $\upmu$s, with a 2-$\upmu$s separation between the two bins. For superposition states $|e\rangle+|l\rangle$, and $|e\rangle+i|l\rangle$, an unbalanced Mach-Zehnder interferometer (MZI) is typically required for analysis. Here, we use the NLPE memory itself as an unbalanced MZI \cite{NSR,RN137} for analysis by splitting the latter $\pi_{13}$ pulse into two $(\frac{\pi}{2})_{13}$ pulses separated by 2 $\upmu$s. The two input pulses resulted in three output pulses, enabling analysis of the superposition states by observing constructive and destructive interference in the central bin of the output. This interference was controlled by applying a phase shift between the two $(\frac{\pi}{2})_{13}$ pulses.
    
    Figure~\ref{qubits}A and Figure~\ref{qubits}B provide photon counting histograms for input qubits of $|e\rangle$, $|l\rangle$ and $|e\rangle+|l\rangle$, with an average input photon number per qubit (in two temporal modes) $\mu_q=1.07$. Histograms for measurements for $|e\rangle+i|l\rangle$ are provided in figure~\ref{smfig2}C. The measured total fidelity $F_T = 89.7 \pm 1.5$\%. Here, $F_T = \frac{1}{3}(F_{|e\rangle} + F_{|l\rangle})/2 + \frac{2}{3}(F_{|e\rangle+|l\rangle} + F_{|e\rangle+i|l\rangle})/2$, where $F_{|i\rangle}$ denotes the storage fidelity of the input qubit $|i\rangle$ \cite{RN153,NSR,RN95}. Additional measurements of storage fidelity are performed for various input photon levels (Table 1). The total fidelity improves to $97.7 \pm 0.8$\% with $\mu_q = 4.21$ and decreases to $86.1 \pm 2.0$\% with $\mu_q = 0.66$. As shown in Figure~\ref{qubits}C, all measured fidelities exceeds the maximal fidelity achievable using a classical ``measure-and-prepare" strategy, considering the finite efficiency and the statistics of input coherent states \cite{RN6,RN153,NSR}. This unambiguously demonstrates that the long-lived integrated quantum memory operates within the quantum regime.
    
    To create a practically competitive device, an intuitive and important target would be beating optical fiber delay line, specifically achieving a storage efficiency greater than the transmission efficiency of a telecom fiber for the same storage duration. The storage efficiency of the current device is $12.0\pm0.5$\% at 1.021 ms, outperforming the transmission efficiency of a fiber delay line by more than three orders of magnitude. The storage time is two orders longer than current photonic quantum memories in integrated devices (see figure S4 for an overview), laying out the foundation for practical applications of integrated quantum memories in large-scale quantum networks \cite{dqc,piqm}.
    % We list the performances on storage efficiency and storage time of integrated quantum memories for light in Figure~\ref{compare}, together with the performance of the fiber delay line as indicated with the blue dashed line.
    
\subsection*{Discussion}
    The short storage time of integrated quantum memories \cite{RN147,RN157,RN61,RN4,piqm} has been a substantial obstacle to their practical applications in long-distance quantum networks \cite{dqc,qr0,dlcz,RN3,oqm,space}. Here, we demonstrate a 1.021-ms integrated quantum memory for photonic qubits, based on a laser-written optical waveguide combined with DD through coplanar electric waveguide. This storage time is sufficient to support the round-trip communication between quantum repeater nodes separated by 100 km \cite{RN3,RN50}. The integration of optical waveguides and coplanar electric waveguides enables precise and efficient manipulations of hyperfine transitions of rare-earth dopants. The resulting performance, including fidelity and efficiency, is competitive compared to similar implementations in bulk crystals \cite{RN170,DD}. This device could support multiplexed operations, primarily in the time and frequency domains, to increase its multimode capacity. While in the spatial domain, the multimode capacity is currently constrained by challenges in confining RF fields. The demonstrated storage efficiency of $12.0\pm0.5$\% surpasses the transmission efficiency of a fiber delay line and further enhancing the efficiency to unity could be achieved by employing impedance-matched optical cavities \cite{cavity,cavityth}. Additionally, the use of a low-concentration sample with small spin inhomogeneous broadening suggests the potential for extending storage times to the order of minutes by operating in critical magnetic fields \cite{RN95, RN47}. Such advancements could unlock new opportunities for transportable quantum memories \cite{space, sat, RN95}.

\subsection*{Materials and Methods}
\begin{figure} % Do NOT use \begin{figure*}
	\centering
	\includegraphics[width=1\textwidth]{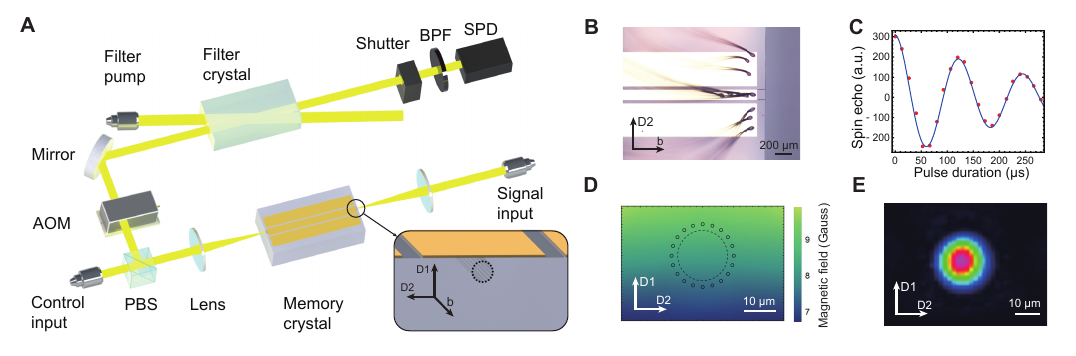} % for an image file named example_figure.*
	% Pick an appropriate width - in print, figures are usually one or two columns wide, which can
	% be approximated by 0.3\textwidth or 0.6\textwidth respectively. Use appropriate label sizes.

	% Captions go below figures
	\caption{\textbf{Experimental setup.}
            (\textbf{A}) Schematic of the setup. The signal input, polarized along D1 axis of $\mathrm{^{151}Eu^{3+}}$:$\mathrm{Y_2SiO_5}$ crystal, is coupled to the optical waveguide fabricated in the memory crystal. After retrieval, the beam is temporally gated by acousto-optic modulators (AOM) and then spectrally filtered by filter crystal and a band pass filter (BPF). A fiber-coupled single photon detector (SPD) finally detects the signal. The control input, polarized along D2 axis of $\mathrm{^{151}Eu^{3+}}$:$\mathrm{Y_2SiO_5}$ crystal, is coupled into the waveguide in the opposite direction and combined with the signal beam through polarization beam splitter (PBS). The control input applies the preparation sequence and optical $\pi$ pulses. The filter crystal is prepared by a pump beam to create a transparent band at signal frequency. Both crystals are housed in the same cryostat, operating at a temperature of 3.2 K. The inset provides a schematic of the input surface of the memory crystal.
            (\textbf{B}) Microscope image of the coplanar electric waveguide fabricated on the top surface of the $\mathrm{^{151}Eu^{3+}}$:$\mathrm{Y_2SiO_5}$ crystal. The waveguide is connected with multiple wires to accommodate high-power RF pulses. The electric waveguide is connected to a RF source and terminates with a 50 ohms load (not shown here).
            (\textbf{C}) Optically detected spin nutation measurement on the hyperfine transiton between $|\pm1/2\rangle_g$ and $|\pm3/2\rangle_g$, achieved with a peak RF power of 4 W, demonstrating a Rabi frequency of 16.7 kHz.
            (\textbf{D}) Simulation of the magnetic field distribution provided by the coplanar electric waveguide. The dashed circle indicates the optical mode within the optical waveguide, while small circles represent fabrication tracks.
            (\textbf{E}) Measured optical mode within the waveguide at the end surface of the crystal.
        }
	\label{setup} % give each figure a logical label name
\end{figure}

    The optical waveguide is fabricated by a femtosecond laser micromachining system (WOPhotonics, Lithuania). The femtosecond laser is set to 1030 nm with 201.9 kHz repetition rate and 210 fs pulse duration. During fabrication the laser is polarized along D2 axis of $\mathrm{^{151}Eu^{3+}}$:$\mathrm{Y_2SiO_5}$ crystal and projects along the D1 axis of $\mathrm{^{151}Eu^{3+}}$:$\mathrm{Y_2SiO_5}$ crystal. The laser is focused by a $\times$100 objective with numerical aperture $\text{NA} =0.7$. The laser moves along b axis to form a notch. We constructed a cylinder with 20 notches with radius $r=20$ $\upmu$m and centered 15 $\upmu$m under the top surface of the crystal. To compensate the power reduction of focus point by depth, the laser power changes gradiently along D1 axis, from 68 nJ for deepest to 64 nJ for shallowest, resulting consistent notches. This symmetrical structure supports all polarization \cite{RN121,RN162}. The free-space coupling efficiency from control input to signal input (as shown in Figure~\ref{setup}A) is 50\% for light polarized along D1 axis and 25\% for polarization along D2 axis respectively. The optical path transmission efficiency without the waveguide is approximately 65\%. We obtain the insertion loss is 1.1 dB for D1 axis polarization and 4.1 dB for D2 axis polarization for the device . The coplanar electric waveguide is manufactured by lift-off technique after the fabrication of optical waveguide. Using ultraviolet lithography (Karl Suss, MABA6) on photoresist (NR9-3000PY) to form a 1 $\upmu$m thick pattern and then we use electron beam evaporation (K.J. Lesker, LAB 18) to coat 50 nm Ti, 720 nm Cu and 30 nm Au onto the crystal to form the electric waveguide.

% If your text is very short you might need to uncomment the following line to avoid
% layout problems with the figures and tables.
\newpage

%%%%%%%%%%%%%%%% MAIN TEXT FIGURES %%%%%%%%%%%%%%%

\begin{figure} 
	\centering
	\includegraphics[width=1\textwidth]{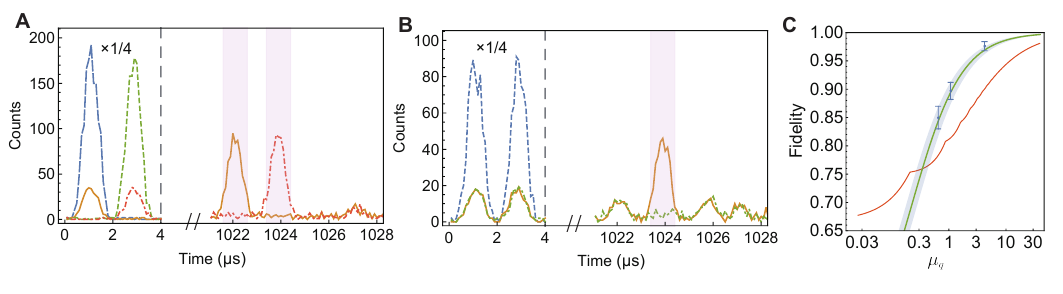} 
	\caption{\textbf{Integrated storage of time-bin qubits for 1.021 ms.}
            (\textbf{A}) Photon-counting histogram for input qubits of $|e\rangle$ and $|l\rangle$, with average input photon number per quibt $\mu_q=1.07$. The inputs are shown with blue ($|e\rangle$) and green dashed lines ($|l\rangle$). The outputs for $|e\rangle$ and $|l\rangle$ are shown as orange solid and red dash-dotted lines, respectively. The light purple shaded area highlights the detection windows, while data on the left of the dashed gray line are scaled by a factor of 1/4 for visualization, for (A) and (B).
            (\textbf{B}) Photon-counting histogram for measurements of $|e\rangle+|l\rangle$ states with $\mu_q=1.07$. The blue dashed line represents the input, while the orange solid and green dash-dotted lines correspond to measurements with constructive and destructive interference, respectively.
            (\textbf{C}) Total fidelity of the memory as a function of the input photon number per qubit $\mu_q$. The red solid line indicates the classical bound, which is the maximum fidelity achievable by any classical device considering the storage efficiency of 12.0\% and the statistics of input coherent states \cite{RN6}. The blue dots represent the measured storage fidelity (see Table 1), while the green line shows the theoretical fidelity calculated based on the measured noise and storage efficiency. Error bars correspond to one standard deviation. Details of the fidelity calculation are provided in the Supplementary Material.
        }
	\label{qubits} 
\end{figure}

%%%%%%%%%%%%%%%% MAIN TEXT TABLES %%%%%%%%%%%%%%%

\begin{table} % Do NOT use \begin{table*}
	\centering
	% Captions go above tables
	\caption{\textbf{Storage fidelity with various input levels.}
	   $\mu_q$ is the average photon number per qubit. $F_{|e\rangle}$ is the fidelity for input qubit $|e\rangle$ and the definition is similar for $F_{|l\rangle}$, $F_{|e\rangle+|l\rangle}$ and $F_{|e\rangle+i|l\rangle}$. $F_T$ is the total fidelity.}
	\label{fidelity_table} % give each table a logical label name
	
	\begin{tabular}{lccccc} % four columns, alignment for each
		\\
		\hline
            $\mu_q$ & 
            $F_{|e\rangle}$ & 
            $F_{|l\rangle}$ & 
            $F_{|e\rangle+|l\rangle}$ & 
            $F_{|e\rangle+i|l\rangle}$ & 
            $F_{T}$\\ 
            \hline
            $0.66$ & 
            $90.7\pm1.3\%$ &
            $90.8\pm1.4\%$ & 
            $84.2\pm2.3\%$ &
            $83.3\pm2.3\%$ &
            $86.1\pm2.0\%$ \\
            
            $1.07$ & 
            $92.4\pm1.0\%$ &
            $93.8\pm0.9\%$ & 
            $87.7\pm1.7\%$ &
            $88.3\pm1.7\%$ &
            $89.7\pm1.5\%$ \\
            
            $4.21$ & 
            $98.6\pm0.5\%$ &
            $98.2\pm0.5\%$ & 
            $97.5\pm0.9\%$ &
            $97.2\pm0.9\%$ &
            $97.7\pm0.8\%$ \\
		\hline
	\end{tabular}
\end{table}

%%%%%%%%%%%%%%%% REFERENCES %%%%%%%%%%%%%%%

\clearpage % Clear all remaining figures and tables then start a new page

% The list of references goes after the main text and before the acknowledgements
% When preparing an initial submission, we recommend you use BibTeX, like this:
%
\bibliography{reference} % for a file named science_template.bib
\bibliographystyle{sciencemag}

% After the paper has completed peer review and been revised ready for acceptance,
% you should comment out the lines above and copy-paste the contents of your .bbl
% file here instead. This will help ensure that our conversion software works correctly.
% Remember to re-run BibTeX first - check the timestamp!
%
% Example of the first three entries copy-pasted from science_template.bbl:
%
%\begin{thebibliography}{1}
%
%\bibitem{example}
%A.~N. {Author}, An example reference. \emph{Journal of Improbable Research}
%  \textbf{1}, 67 (2020).
%
%\bibitem{example2}
%F.~M. {Surname}, S.~{Author}, A second example. \emph{Interesting Research
%  Letters} \textbf{32}, 897 (2019).
%
%\bibitem{example_preprint}
%P.~{One}, P.~{Two}, P.~{Three}, {An unpublished preprint}. \emph{preprint}
%  (2021), arXiv:2101.12345.
%
%\end{thebibliography}

%%%%%%%%%%%%%%%% ACKNOWLEDGEMENTS %%%%%%%%%%%%%%%

\section*{Acknowledgments}
    \paragraph*{Funding:}
        This work is supported by the Innovation Program for Quantum Science and Technology (No. 2021ZD0301200), the National Natural Science Foundation of China (Nos. 12222411, 11821404 and 12204463). The fabrication process is partially carried out at the USTC Center for Micro and Nanoscale Research and Fabrication. Z.-Q.~Z. acknowledges the support from the Youth Innovation Promotion Association CAS.
    \paragraph*{Author contributions:}
        Z.-Q.~Z. designed and supervised the experiment. Y.-P.~L. implemented the experiment and analyzed the data with the help from Z.-W.~O., T.-X~Z., C.~L. and G.-C.~G.. Y.-P.~L. fabricated the sample with the help of Z.-W.~O., M.-X.~S. and Y.-J. H.. Y.-P.~L. and Z.-Q.~Z. wrote the paper with the help of others. Z.-Q.~Z. and C.-F.~L. supervised the project. All authors discussed the experimental procedures and results.
    \paragraph*{Competing interests:}
        There are no competing interests to declare.
    \paragraph*{Data and materials availability:}
        All data needed to evaluate the conclusions in the paper are present in the paper and/or the Supplementary Materials.

\subsection*{Supplementary materials}
Supplementary Text\\
Figures. S1 to S5\\
References %\textit{(7-\arabic{enumiv})} % automatically fills out the last reference number
% (filling out the other numbers automatically is possible but fiddly and liable to break)

%%%%%%%%%%%%%%%% END OF MAIN TEXT %%%%%%%%%%%%%%%

\newpage

%%%%%%%%%%%%%%%% START OF SUPPLEMENT %%%%%%%%%%%%%%%

% Figures, tables, equations and pages in the supplement are numbered S1, S2 etc.
\renewcommand{\thefigure}{S\arabic{figure}}
\renewcommand{\thetable}{S\arabic{table}}
\renewcommand{\theequation}{S\arabic{equation}}
\renewcommand{\thepage}{S\arabic{page}}
\setcounter{figure}{0}
\setcounter{table}{0}
\setcounter{equation}{0}
\setcounter{page}{1} % not 0 as \newpage already started a supplementary page
% References continue the numbering from the main text.

%%%%%%%%%%%%%%%% SUPPLEMENT TITLE PAGE %%%%%%%%%%%%%%%

\begin{center}
\section*{Supplementary Materials for\\ \scititle}

% Author list for the supplement
% Indicate the corresponding authors, but do NOT include institutions here
% It would be nice if the template auto-generated this, but doing so is complicated...
	Yu-Ping~Liu$^{1,2,3}$, %\ast\dagger
	Zhong-Wen~Ou$^{1,2,3}$,
	Tian-Xiang~Zhu$^{1,2,3}$,  \\
	Ming-Xu~Su$^{1,2,3}$, 
	Chao~Liu$^{1,2,3}$,
	Yong-Jian Han$^{1,2,3,4}$, \\
 Zong-Quan~Zhou$^{\mathrm{a},1,2,3,4}$, 
	Chuan-Feng~Li$^{\mathrm{b},1,2,3,4}$, 
	Guang-Can~Guo$^{1,2,3,4}$\\ % we're not in a \author{} environment this time, so use \\ for a new line
        \small$^\mathrm{a}$zq\_zhou@ustc.edu.cn\\
        \small$^\mathrm{b}$cfli@ustc.edu.cn\\
\end{center}

% Fill out the numbers for each type of supplementary material,
% and delete any lines that aren't applicable.
% These are just example numbers that don't match the rest of this template.
\subsubsection*{This PDF file includes:}
Supplementary Text\\
Figures S1 to S5\\
Reference

\newpage

%%%%%%%%%%%%%%%% MATERIALS AND METHODS %%%%%%%%%%%%%%%

%reference  \cite{methods}.
%Fig.~\ref{fig:example}, %Table~\ref{tab:example},
%Fig.~\ref{fig:sup_example} and Table~\ref{tab:sup_example}.
%Cite references in the usual way \cite{example2},
%including any that are only cited in the supplement \cite{sm_example,conference_example}.

%The numbering of figures, tables, equations and pages has been reset to start from S1, as in
\textbf{Supplementary Text}

\subsubsection*{Additional details about the storage device}
\begin{figure} % Do not use \begin{figure*}
	\centering
	\includegraphics[width=0.65\textwidth]{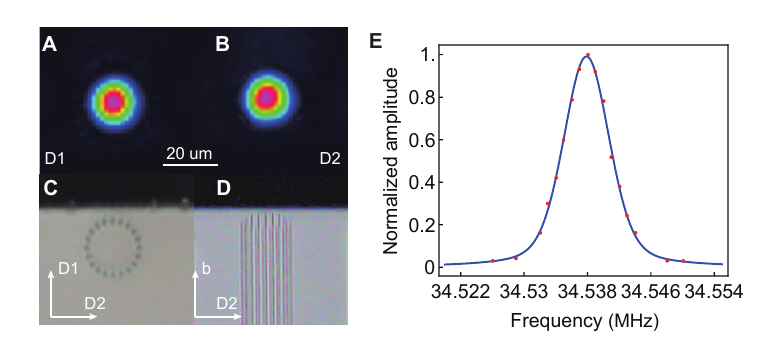} 
	\caption{\textbf{Device properties.}
            (\textbf{A}) and (\textbf{B}) The guide mode profiles for light polarized along the D1 axis (A) and the D2 axis (B). The mode field diameters are 16.3 $\upmu$m $\times$  16.3 $\upmu$m (A) and 16.2 $\upmu$m $\times$  16.5 $\upmu$m (B).
            (\textbf{C}) and (\textbf{D}) The front view and the top view of the optical waveguide.
            (\textbf{E}) The inhomogeneous broadening of the hyperfine transition of Eu$^{3+}$ ions inside the optical waveguide. The red dots represent experimental data acquired by optically detected spin nutation measurement.The blue line is the fitted curve using Voigt distribution. The full width at half maximum bandwidth of the fitted curve is $7.8\pm0.1$~kHz. 
            }
	\label{smfig1} 
\end{figure}

\begin{figure} % Do not use \begin{figure*}
	\centering
	\includegraphics[width=0.8\textwidth]{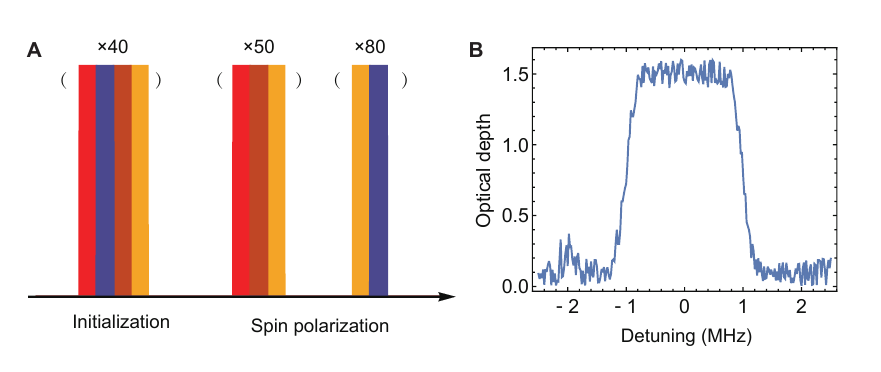} 
	\caption{\textbf{Spectrum preparation.}
            (\textbf{A}) The preparation sequence consists of a series of 1-ms pump processes, with each pulse corresponding to a specific transition. The red pump pulse corresponds to the $|\pm1/2\rangle_g \rightarrow |\pm5/2\rangle_e$ transition, blue represents $|\pm5/2\rangle_g \rightarrow |\pm3/2\rangle_e$, brown represents $|\pm1/2\rangle_g \rightarrow |\pm3/2\rangle_e$ and yellow represents $|\pm3/2\rangle_g \rightarrow |\pm5/2\rangle_e$. 
            (\textbf{B}) The optical absorption profile as prepared at the $|\pm1/2\rangle_g \rightarrow |\pm5/2\rangle_e$ transition.
            }
	\label{smfig15} 
\end{figure}

The optical waveguide constrains light through refractive index change around notches (Figure~\ref{smfig1}). The fabrication process also induces lattice structure deformation, which results in a slightly larger inhomogeneous broadening of optical transitions of Eu$^{3+}$ ions in waveguide than that in the bulk crystal \cite{inhomo, NSR}. Here we measured an optical inhomogeneous broadening of 0.80 GHz and 0.65 GHz for Eu$^{3+}$ ions inside and outside the waveguide, respectively. The peak absorption depth is 2.09 for Eu$^{3+}$ ions inside waveguide. The spectral feature is prepared by spectral hole burning technique (Figure~\ref{smfig15}A), see \cite{NSR} for detail. The prepared 1.8-MHz optical absorption profile is shown in Figure~\ref{smfig15}B.

The efficiency of NLPE memory can be modeled by \cite{RN137,NSR}:
\begin{equation}
    \eta=d^2 e^{-d}\left(\eta_{\text {control }}\right)^4 e^{-\gamma_{13}{ }^2 t_{31}{ }^2 /\left(2 \ln (2) / \pi^2\right)} e^{-\gamma_{\bar{3} \bar{5}}{ }^2 t_{42}{ }^2 /\left(2 \ln (2) / \pi^2\right)} e^{-2 \gamma \cdot t_{42}},
    \label{eq_efficiency} 
\end{equation}
where $\gamma_{13}$ is the inhomogeneous broadening of transition $|\pm1/2\rangle_g\rightarrow |\pm3/2\rangle_g$, $\gamma_{\bar{3}\bar{5}}$ is the inhomogeneous broadening of transition $|3 / 2\rangle_{\mathrm{e}} \rightarrow|5 / 2\rangle_{\mathrm{e}}$ and $\gamma$ is the effective optical decoherence rate. $\eta_{\text{control}}$ is the average transfer efficiency of optical $\pi$ pulses. $t_{31}$ is the interval between the first and third optical $\pi$ pulses, while $t_{42}$ is the interval between the second and fourth optical $\pi$ pulses. By measuring the echo amplitude depending on variable $t_{31}$ and $t_{42}$ (Figure~\ref{smfig2}A, B), we fit the parameters and get: $\gamma_{13}=6.0\pm0.9$ kHz, $\gamma_{\bar{3}\bar{5}}=18\pm1$ kHz, $\gamma=8\pm4 $ kHz, $\eta_{\text{control }}=85$\% \cite{RN137,NSR}. The value of $\gamma_{13}$ is approximately consistent with the direct measurement on spin inhomogeneous broadening using optically detected spin resonance (Figure~\ref{smfig1}E).

\begin{figure} 
	\centering
	\includegraphics[width=1\textwidth]{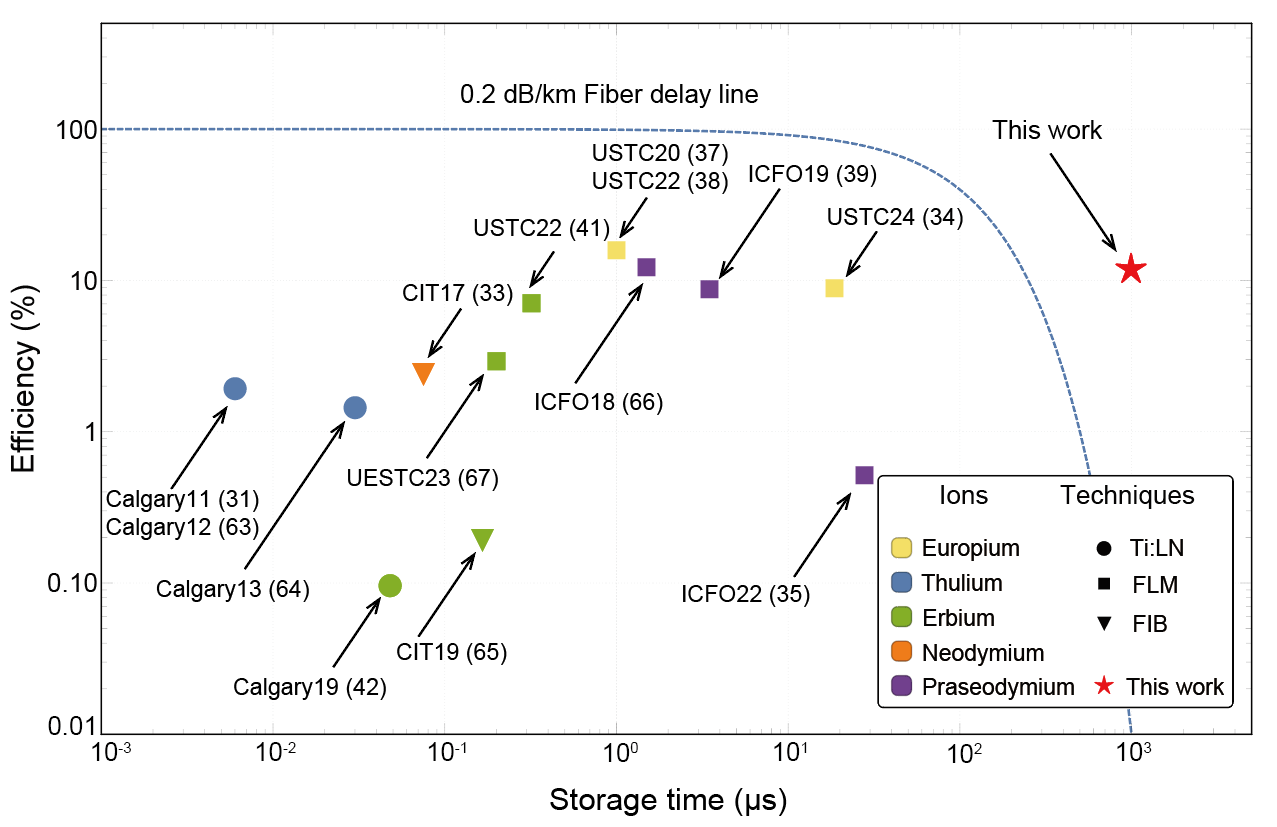} 
	\caption{\textbf{The performances of integrated quantum memories for light.}
            The colors represent various atomic elements and the shape of dots represent various fabrication techniques, as indicated in the inset. Circular dots are Titanium-indiffused Lithium Niobate (Ti:LN) waveguides \cite{RN4,ln1,ln2,ln3}, triangular dots are focused ion beam (FIB) milling cavities \cite{RN147,f2}, square dots are waveguides fabricated by femtosecond laser micromachining (FLM) \cite{NSR,RN162,m2,RN61,m3,RN195,RN157,erln23}. The blue dashed line represents the performance of a telecom fiber delay line with a loss coefficient of 0.2 dB/km. The red star marks the current device, which is the only integrated quantum memory to outperform the fiber delay line. The data presented corresponds to the longest storage times directly demonstrated for single-photon-level quantum memory in each experiment. It is worth noting that long-duration quantum memories have also been achieved in bulk crystals \cite{DD,RN170}, though these are not included in this figure.
        }
	\label{compare} 
\end{figure}

The quantum storage time of this device is two orders of magnitude longer than that of previous integrated quantum memories. In Figure~\ref{compare}, we present the storage efficiency and storage time of integrated quantum memories for light, alongside the performance of a fiber delay line, as indicated by the blue dashed line.

\subsubsection*{The storage fidelity of time-bin qubits}
The fidelity of time-bin qubits is analyzed by projection on the ideal state and the orthogonal state, with photon counts of $N_\mathrm{+}$ and $N_\mathrm{-}$, respectively. The fidelity is then given by
\begin{equation}
\label{eq1}
    F=\frac{N_\mathrm{+}}{N_\mathrm{+}+N_\mathrm{-}}.
\end{equation}

For $\mu_q=1.07$, the measurement results for $|e\rangle$, $|l\rangle$, $|e\rangle+|l\rangle$ are provided in Figure~\ref{qubits}A,B. Additionally, Figure~\ref{smfig2}C shows the photon-counting histogram for input qubits of $|e\rangle+i|l\rangle$. 

The measured counts should include both the echo signal and the unconditional noise:
\begin{align}
    N_\mathrm{+}&=(\mu_q\eta_{M}F_c+p_n)N;\\
    N_\mathrm{-}&=(\mu_q\eta_{M}(1-F_c)+p_n)N,
\end{align}
where $\eta_M$ is memory efficiency, $p_n$ is unconditional noise probability and $N$ is experiment repetition. $F_c$ is classical fidelity which we measured to be 99.7 \%. Here we assume $F_c=1$ for simplicity, which also gives a stricter upper limit on fidelity bound. By substituting $N_\mathrm{+}$ and $N_\mathrm{-}$ into Equation~\ref{eq1}, the theoretical storage fidelity can be predicted as:
\begin{equation}
    F(\mu_q)=\frac{F_c+p_n/\mu_q\eta_{M}}{1+2p_n/\mu_q\eta_{M}},
\end{equation}
which is indicated as the green line in Figure~\ref{qubits}C.

According to reference \cite{RN6,RN100}, a classical bound on storage fidelity for weak coherent states can be derived. Since the coherent state has a Poisson distribution of $P(\mu_q, n)=\mathrm{e}^{-\mu_q} {\mu_q^n}/{n}$ with average photon number $\mu_q$, it's possible an intercept-resend attack \cite{attack} can gain additional information from multi-photon incidents. The attacker can measure and then prepare a state when there's high enough photon incident $n\geqslant n_\mathrm{min}$ from the input, where $n_\mathrm{min}$ is the minimal $i$ that matches $\left(1-P\left(\mu_q, 0\right)\right) \eta_M-\sum_{n>i+1} P\left(\mu_q, n\right)\geq 0$. And the fidelity bond is:
\begin{equation}
    F_{\text {classical}}(\mu_q)=\dfrac{\left(\dfrac{n_{\min }+1}{n_{\min }+2}\right) \mathit{\Gamma}+
    \displaystyle\sum\limits_{n \geqslant  n_{\min }+1}{\dfrac{n+1}{n+2} P(\mu_q, n)}}
    {\mathit{\Gamma}+\displaystyle\sum\limits_{n \geqslant n_{\min}+1} P(\mu_q, n)},
\end{equation}
where $\mathit\Gamma=\left(1-P\left(\mu_q, 0\right)\right) \eta_M-\sum_{N>N_\mathrm{min}+1} P\left(\mu_q, n\right)$. This limit is shown with red solid line in Figure~\ref{qubits}C.

\subsubsection*{Characterization of the coplanar electric waveguide}
%The coplanar electric waveguide exhibits a decent response as mentioned in main text. The high transmission rate ensures that the RF signal is applied effectively, while the low reflection rate minimizes unintended RF fields \cite{cpw2}. 

We have also measured the performance of coplanar electric waveguide fabricated on the D1 $\times$ b plane. With the same input RF power, the Rabi frequency was 0.8 times that of the current D2 $\times$ b configuration. Therefore, we select the latter configuration for efficient spin manipulation.

\begin{figure} % Do not use \begin{figure*}
	\centering
	\includegraphics[width=1\textwidth]{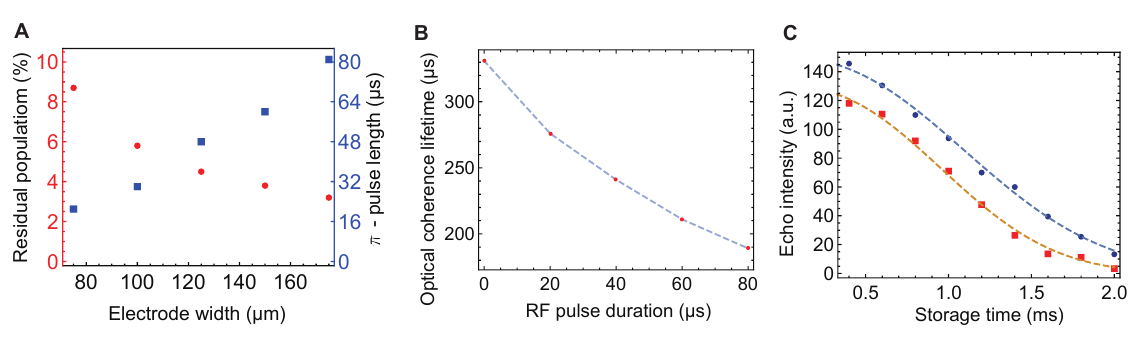} 
	\caption{\textbf{Characterization of coplanar electric waveguide}
            (\textbf{A}) Residual population (red circular dots) per CHS XX sequence and the length of $\pi$ pulses (blue squares) determined by spin nutation measurements for variable electrode width. The peak power of the RF pulses is fixed at 4 W.
            (\textbf{B}) Optical coherence lifetimes of Eu$^{3+}$ measured immediately after a RF pulse excitation with variable pulse length. The peak power of the RF pulses is fixed at 6 W and the frequency is 10 MHz.
            (\textbf{C}) The lifetime of NLPE-DD memory. The red squares are data with a XXXX sequence and the blue circulars are data with a XY4 sequence. The lines are the corresponding fitted decay curve.
            }
	\label{smfig3} 
\end{figure}

The RF field intensity correlates with the width of central electrode. We measure the $\pi$ pulse length with a fixed RF pulse peak power of 4~W using spin nutation measurement (Figure~\ref{smfig3}A). For all other measurements, we use complex hyperbolic secant (CHS) pulses to generate more efficient and robust $\pi$ pulses \cite{chs1,chs2}.

Smaller electrodes could generate stronger magnetic fields, but the field homogeneity would degrade within the optical waveguide, leading to errors in $\pi$ pulses. The size of the optical waveguide is chosen to minimize damage to the crystal and maintain a small inhomogeneous broadening. A smaller optical waveguide could achieve better RF field homogeneity, but the increased optical and spin inhomogeneous broadening could decrease NLPE efficiency. Therefore, after determining appropriate optical waveguide size, we optimize the width of the coplanar electric waveguide to achieve a balance between RF field strength and homogeneity.

To benchmark the performance of CHS $\pi$ pulses for different electric waveguide widths, we detect the residual population through absorption measurements. Since the XY4 sequence is robust to pulse errors, the pulse error is estimated with a set of XX sequences similar to those in Ref. \cite{DD}. We measure the population error after 1 to 6 XX sequences and fit for the population error per XX sequence, the resulting residual population per XX pulse for different electrode widths is shown in Figure~\ref{smfig3}A. A width of 150 $\upmu$m is sufficient to achieve the required population transfer efficiency. The residual population in $|\pm3/2\rangle_g$ tested for the electrode is $3.8\pm0.1\%$ after an XX sequence, which corresponds to a single-pulse error of $(0.062\pm0.001)\pi$ \cite{dderror}. Combined with XY4 sequence, this electric waveguide provides a satisfactory signal-to-noise ratio.

 The CHS $\pi$ pulse employed in actual DD has a length of 60~$\upmu$s and a peak power of 4~W. As the electrodes are positioned closer to the optical waveguide, the heating effects should be evaluated. We measure the optical coherence lifetimes of Eu$^{3+}$ ions, which is sensitive to temperature \cite{temp}, immediately after an RF pulse excitation. Here, the RF frequency is set to 10 MHz, far off-resonance to the Eu$^{3+}$ ions to avoid direct interaction between the RF field and the the Eu$^{3+}$ ions and the input peak power is raised to 6~W. According to the data shown in Figure~\ref{smfig3}B, the optical coherence lifetime is still longer than 210 $\upmu$s after a 60 $\upmu$s RF pulse. Taking the optical evolution time in NLPE memory into consideration, the efficiency drop due to heating effect is less than 3\%. In addition, the evolution in the optical transition starts at 0.1 ms after the last CHS $\pi$ pulse in NLPE-DD memory, so the device is further cooled during this interval and we deduce that the efficiency loss due to RF heating is negligible in our experiments. %comparing with the optical coherence time of 330$\upmu$s with no RF pulse.

We have compared the performance of XXXX and XY4 DD sequences. The $1/e$ lifetime of NLPE-DD memory is 1.6 ms for XXXX sequence and 1.9 ms for XY4 sequence, respectively (Figure~\ref{smfig3}C). In addition, XY4 sequence provides an overall higher efficiency so it is employed in the quantum storage experiments.

\begin{figure} % Do not use \begin{figure*}
	\centering
	\includegraphics[width=1\textwidth]{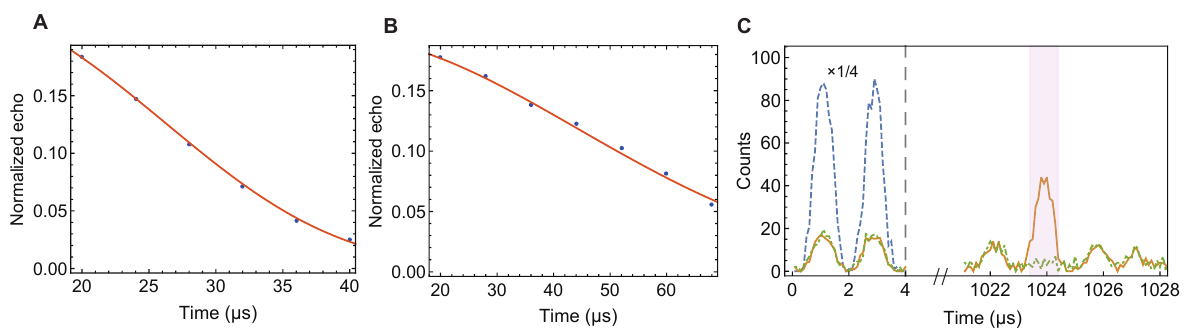} 
	\caption{\textbf{NLPE memory details.}
            (\textbf{A}) and (\textbf{B}) Decay of NLPE echo by delaying the time interval between two $\pi_{13}$ (A) and two $\pi_{35}$ (B). The blue dots are normalized echo area and the red lines are fits based on Equation~\ref{eq_efficiency}.
            (\textbf{C}) Photon-counting histogram for analysis of $|e\rangle+i|l\rangle$ qubit with $\mu_q=1.07$. The inputs are shown with blue dashed line. The orange solid line and the green dash-dotted line represent the measurements with constructive and destructive interference, respectively. The light purple shaded area indicates the detection windows and the data on the left side of dashed gray line are scaled by a factor of 1/4 for visual effects.
            }
	\label{smfig2} 
\end{figure}

Meanwhile XY4 sequence provides a smaller residual population according to Ref. \cite{dderror}. The absorption difference is too small to be detected directly in the current system. Therefore, instead of directly measuring the absorption, we estimate the error of XY4 sequence by comparing the efficiency with and without applying the sequence. Since the efficiency decays with time at different rate with and without DD, we compare the fitted intercept at zero spin-wave evolving time. From the fitting in Figure~\ref{smfig3}C, the NLPE-DD memory with XY4 sequence has a zero-time intercept efficiency of $19.79\pm 0.03$\%, which corresponds to the maximum NLPE-DD efficiency with zero spin-wave storage time. Meanwhile, for NLPE memory of spin wave evolution without DD as shown in Figure~\ref{smfig2}B, the zero-time intercept efficiency is $19.85\pm 0.04$\%. Based on these two zero-time intercept efficiencies, we estimate that our device has a $99.7\pm0.2\%$ rephasing efficiency for spin transitions, which puts a upper bound for residual population of $0.3\pm0.2$\% after an XY4 sequence.

%Since XY4 sequence is not sensible with pulse error, we test the pulse imperfection by measuring residual population in $|\pm3/2\rangle_g$ after a series of XX sequences and get a population error of $3.8\pm0.1\%$ per XX sequence, which corresponds to a pulse error of $(0.062\pm0.001)\pi$\cite{DD}. (XY4 0.3\% calculation)

%%%%%%%%%%%%%%%% SUPPLEMENTARY TEXT %%%%%%%%%%%%%%%

%\subsection*{Supplementary Text}
%The Supplementary Text section can only be used to directly support statements made in the main text e.g. to present more detailed justifications of assumptions, investigate alternative scenarios, provide extended acknowledgements etc. Material in this section cannot claim results or conclusions that weren't mentioned in the main text. To refer to this section from the main text, just write (Supplementary Text).

%\subsubsection*{Example supplement heading}

%The two main sections of the supplement can be split up using headings.

% If your supplement is very short you might need to uncomment the following line to avoid
% layout problems with the figures and tables.
\newpage

%%%%%%%%%%%%%%%% SUPPLEMENTARY FIGURES %%%%%%%%%%%%%%%

%%%%%%%%%%%%%%%% SUPPLEMENTARY TABLES %%%%%%%%%%%%%%%

%%%%%%%%%%% CAPTIONS FOR OTHER SUPPLEMENTARY FILES %%%%%%%%%%

%\clearpage % Clear all remaining figures and tables then start a new page

%\paragraph{Caption for Movie S1.}
%\textbf{All captions must start with a short bold sentence, acting as a title.} Then explain what is shown in the supplementary video file. Give as much detail as you would for a figure e.g. explain axes, color maps etc. If the video is an animated equivalent of one of the static figures, state e.g.%Animated version of Figure~\ref{fig:example}.'

%\paragraph{Caption for Data S1.}
%\textbf{All captions must start with a short bold sentence, acting as a title.} Then explain what is included in the supplementary data file. Give as much detail as you would for a table e.g. explain the meaning of every column, units used, any special notation etc.

%%%%%%%%%%%%%%%% SUPPLEMENTARY REFERENCES %%%%%%%%%%%%%%%

% Do NOT include a reference list in the supplement.
% All references must be in a single list at the end of the main text.
% The copyeditors will ensure that the correct reference list appears with each version of the paper
% (print, HTML, PDF, mobile app, metadata for bibliographic databases etc.)

\end{document}